# Expansive Open Fermi Arcs and Connectivity Changes Induced by Infrared Phonons in ZrTe$_5$

Lin-Lin Wang

Ames Laboratory, U.S. Department of Energy, Ames, IA 50011, USA

## Abstract

Expansive open Fermi arcs covering most of the surface Brillouin zone (SBZ) are desirable for detection and control of many topological phenomena, but so far has been only reported for Kramers-Weyl points, or unconventional chiral fermions, pinned at time-reversal invariant momentum in chiral materials. Here using first-principles band structure calculations, we show that for conventional Weyl points in ZrTe$_5$ with the chirality of +1/–1 near the BZ center at general momentum induced by one of the infrared phonons, the second lowest $B_{1u}$ mode for breaking inversion symmetry, they can also form expansive open Fermi arcs across the SBZ boundary to occupy most of the SBZ when projected on (001) surface. We reveal that such expansive open Fermi arcs are evolved from the topological surface states that connect multiple surface Dirac points on the (001) surface of the topological insulator phases without lattice distortion in ZrTe$_5$. Furthermore, we find that the connectivity of the induced open Fermi arcs can be changed by the magnitude of the lattice distortion of this infrared phonon mode. Thus, we propose that using coherent optical phonon to modulate lattice parameters can offer ways to induce novel topological features including expansive open Fermi arcs and dynamically control Fermi arcs connectivity in ZrTe$_5$.

llw@ameslab.gov

# I. Introduction

Weyl fermions[1] can emerge in systems where time-reversal symmetry (TRS) or inversion symmetry (IS) or both are broken. Weyl points (WPs) of opposite chirality in momentum space are monopoles (sinks or sources) for Berry[2] curvature. The projection of WPs on a surface without overlap are connected via open-ended Fermi arcs. Weyl fermions give arise to novel topological properties, such as chiral anomaly[3-5], chiral magnetoelectric effect[6, 7], anomalous Hall effect[8], circular photo-galvanic effect[9-12], Weyl orbits[13-15] and nonlocal transport[16]. The WPs at general momentum of Brillouin zone (BZ) formed from two-band crossing with the chirality of +1/–1 are conventional WPs and appear in pairs. Recent studies[17-20] have extended the chiral fermions to unconventional cases or Kramers-Weyl points that are pinned at time-reversal invariant momentum (TRIM) with high degeneracy and chirality in chiral materials, such as RhSi and CoSi. The corresponding open Fermi arcs are expansive and can cover the whole surface Brillouin zone (SBZ). In contrast, because conventional WPs are split from either Dirac points by breaking TRS [21, 22] or the same band inversion by breaking IS[23], the distance among conventional WPs and the corresponding open Fermi arcs are usually small compared to the size of SBZ.

However, here we show that for the conventional WPs emerging near the $\Gamma$ point in $ZrTe_5$ induced by one of the infrared (IR) phonon, the second lowest $B_{1u}$ mode, to break IS and transform the strong or weak topological insulator (TI) phase into a Weyl semimetal (WSM), the open Fermi arcs can also be expansive on $ZrTe_5$ (001) surface to cross BZ boundary and span most of the SBZ. We further reveal that such expansive behavior arises from the topological surface states connecting the multiple surface Dirac points (SDPs) on (001) for the TI phases of $ZrTe_5$ without lattice distortion. This shows that surface terminations of a TI with multiple SDPs[24] can be used as the platform to induce expansive open Fermi arcs, in addition to Kramers-Weyl in chiral materials[17-20]. Expansive open Fermi arcs have the advantage for easy detection and versatile manipulation to realize many topological phenomena, such as spin texture, Weyl orbits[13-15] and nonlocal transport[16]. The connectivity of open Fermi arcs from WP projections, either within the SBZ (intra-SBZ) or crossing the SBZ boundary (inter-SBZ), depends on the detailed surface band structures[25]. Here we show that the connectivity of open Fermi arcs on $ZrTe_5$ (001) surface can be dynamically controlled by the magnitude of the lattice distortion of this $B_{1u}$ phonon mode. On the same surface termination of ZrTe$_5$ (001), this $B_{1u}$ mode can drive the open Fermi arcs connectivity from intra-SBZ to inter-SBZ.

ZrTe$_5$ has attracted intensive interest because it is near the phase boundary between a strong TI (STI) and a weak TI (WTI). A single layer of ZrTe$_5$ has been predicted[26] to be a quantum spin Hall insulator, while the 3D ZrTe$_5$ crystal is either a STI or WTI depending on the small changes in structural parameters used in the band structure calculations. The reason for such behavior is from the band inversion between $p$ orbitals of two different Te sites at $\Gamma$ point, which is very sensitive to the small change in atomic positions. Different experiments have also provided a range of evidence indicating the scenarios from Dirac semimetal (DSM)[27] to TIs[28] to WSM[29]. A recent study[30] has shown that the different properties can be due to different concentration of defects from different growth methods. In high magnetic field, the WPs induced by breaking TRS have also been proposed to explain the anomalous Hall conductivity[29]. Such sensitive dependence of topological states on the small structural change in ZrTe$_5$ offers the opportunity for controlling topological phase transitions[31]. Temperature[32], pressure, volume[33] and strain[34] have been shown to drive the WTI-STI transition with a critical DSM in between for ZrTe$_5$. Thus, lattice distortion from optical phonon modes can be used very effectively to induce topological phase transition in ZrTe$_5$, such as WSM, which has been proposed in other systems[23] where IS is removed by IR phonon modes. Recently in a Terahertz (THz) optical pump-probe experiment, we have induced the STI-DSM-WTI topological phase transition in ZrTe$_5$ by a coherent Raman phonon[35] mode for retaining the IS.

In this paper, we first report the calculated phonons for ZrTe$_5$ from density functional theory[36, 37] (DFT) and compare the zone-center modes to experimental data. Next, we show that the IR phonon, namely two lowest $B_{1u}$ modes, can induce two pairs of WPs around $\Gamma$ point near the Fermi energy ($E_F$) by breaking IS. More importantly, for the second lowest $B_{1u}$ mode, these induced conventional WPs with large enough lattice distortion can change Fermi arcs connectivity and have expansive open Fermi arcs when projected on (001) surface. Then we reveal that such expansive open Fermi arcs arise from the topological surface states connecting the multiple surface Dirac points (SDPs) on (001) for the TIs without lattice distortion. Thus, we propose that surface terminations with multiple SDPs connected by topological surface states can be used as platforms to induce expansive open Fermi arcs and dynamically change Fermi arcs connectivity by coherent optical phonon to modulate lattice parameters.

## II. Computational Methods

Phonons of ZrTe$_5$ have been calculated with density functional perturbative theory[38] (DFPT) using PBE[39] exchange-correlation (XC) functional with D2 parameters[40] for van der Waals (vdW) interaction in DFT[36, 37] as implemented in Quantum Espresso[41] (QE). We use a $\Gamma$-centered Monkhorst-Pack[42] (9×9×5) *k*-point mesh for the primitive ZrTe$_5$ base-centered orthorhombic cell of 12 atoms. For the electronic band structure calculations, the atomic displacements from the chosen optical phonon mode are applied to the experimental lattice parameters[43] to study the phonon induced topological phase transitions. Spin-orbit coupling (SOC) effect is included in all band structure calculations, which are performed in Vienna Ab initio Simulation Package[44, 45] (VASP) with a plane-wave basis set and projector augmented wave (PAW) method[46]. The kinetic energy cutoff is 816 eV for norm-conversing pseudo-potentials in QE and 230 eV for PAW in VASP. The convergence with respect to *k*-point mesh has been carefully checked, with total energy converged below 1 *m*eV/cell. For ionic relaxation, the absolute magnitude of force on each atom is reduced below 0.005 eV/Å. The relaxation in VASP is with an increased kinetic energy cutoff of 287 eV. For phonon calculations, the small displacement method in VASP and Phonopy[47] have also been used. A (3×3×2) q-point mesh is used with the primitive unit cell for DFPT calculation and a (4×2×2) supercell of 384 atoms built with the conventional unit cell is used for the small displacement method. To calculate topological properties, a tight-binding (TB) model based on maximally localized Wannier functions (MLWFs)[48, 49] has been constructed to reproduce closely the band structure including SOC in the range of $E_F \pm 1$eV. Then the spectral functions and Fermi surface (FS) of a semi-infinite surface are calculated using the surface Green's function methods[50-53] as implemented in WannierTools[54].

## III. Results and Discussion

### A. Crystal structure and phonons

The crystal structure of ZrTe$_5$ (Fig.1(a)) has a base-centered orthorhombic unit cell in space group 63 (*Cmcm*). It has a layered structure along *b*-axis with Zr and apical Te (Te$_a$) occupying the 4$c$ site, while dimer Te (Te$_d$) and zigzag Te (Te$_z$) occupying the 8$f$ site. The base center is on the $ab$ plane. The crystal structure can also be seen as a particular arrangement of ZrTe$_8$ clusters, which includes Zr and the neighboring Te atoms (two Te$_a$, four Te$_d$ and two Te$_z$). Along $a$-axis, the cluster by itself is quite close-packed. Then along $c$-axis, the oppositely aligned clusters form sheets in sharing a zig-zag Te chain at the boundaries. Lastly, the ZrTe$_5$ sheets are loosely stacked along $b$-axis with vdW interaction. So the natural cleaving plane gives (010) surface. The next easy to cleave is (001) surface breaking among the zig-zag Te$_z$ chain. These two surfaces and the corresponding 2D SBZs in relation to the bulk BZ are shown in Fig.1(b). The crystal structure has two mirror planes, one normal to the $a$-axis and other to $c$-axis, and also a glide plane normal to the $b$-axis.

The calculated phonon band structure and density of states (DOS) are plotted in Fig.1(c) and (d), respectively. There are two types of phonons showing small or large band dispersion. The phonons with small band dispersion correspond to intra-cluster modes and have sharp peaks in DOS, mostly in the higher energy range beyond 70 $cm^{-1}$. The phonons with large band dispersion correspond to mostly relative motions of the ZrTe$_8$ clusters and have broad peaks in DOS below 70 $cm^{-1}$. The two inversion centers of the crystal structure are not inside ZrTe$_8$ clusters, but with one being between the clusters and the other at the mid-point of Te$_z$-Te$_z$ bond, as shown by the black circles in Fig.1(e) and (f). This means some of the low-energy IR modes can be effective in breaking the IS. Additional phonon calculations using the small displacement method and also PBEsol[55], a different XC functional, are shown in the Appendix to give good agreement.

With the PBE[39] XC functional and also D2 parameters[40] for vdW interaction, the dimensions of the relaxed unit cell are $a$=3.985 Å, $b$=14.393 Å and $c$=13.629 Å, which agree very well to the experimental data[43] (within 1%) of $a$=3.987 Å, $b$=14.502 Å and $c$=13.727 Å. The low-energy zone-center phonon modes are listed in Table 1. The PBE-D2 XC functional gives better agreement with experimental data[56] than local density approximation[57, 58] (LDA), which over-binds, gives smaller lattice parameters for the fully relaxed unit cells and overestimates the phonon frequencies in most cases. Among the listed optical phonon modes, the $A_{1g}$ Raman mode of 37.6 $cm^{-1}$ is the one that we have studied to induce WTI-DSM-STI topological phase transition[35]. To study the effect of breaking IS on topological band structure to induce WPs, we choose the two lowest $B_{1u}$ IR mode of 25.7 and 44.1 $cm^{-1}$ at zone center. The first lowest $B_{1u}$ mode as depicted in Fig.1(e) mostly involves the torsional movement of the ZrTe$_8$ clusters around the $a$-axis. The second lowest $B_{1u}$ mode mostly involves the atomic displacements of Te$_a$ toward zig-zag Te$_z$ chain along $c$-axis as shown in Fig.1(f). We introduce the parameter of $\lambda$ to describe the magnitude of the lattice distortion. For the first lowest $B_{1u}$ mode of 25.7 $cm^{-1}$ with $\lambda$=1.0, it corresponds to the lattice distortion of 0.005 Å for Zr, 0.025 Å for Te$_a$, 0.030 Å for Te$_d$ and 0.035 Å for Te$_z$, respectively. For the second lowest $B_{1u}$ mode of 44.1 $cm^{-1}$ with $\lambda$=1.0, it involves the lattice distortion of 0.014 Å for Zr, 0.056 Å for Te$_a$, 0.015 Å for Te$_d$ and 0.025 Å for Te$_z$, respectively. Notably for the second lowest $B_{1u}$ mode, there is a large displacement of Te$_a$ along the $c$-axis in perpendicular to the (001) surface, which can induce a much larger change in the topological surface states than the first lowest $B_{1u}$ mode. In the following, we will focus on this second lowest $B_{1u}$ mode that induces expansive open Fermi arcs on (001) surface and then also discuss the first lowest $B_{1u}$ mode for comparison.

## B. Bulk band structures and induced Weyl points

The electronic band structures of ZrTe$_5$ with the second lowest $B_{1u}$ lattice distortion are presented in Fig.2. For reference, the band structure of ZrTe$_5$ without lattice distortion is plotted in Fig.2(a). The small flat piece along $\Gamma$-$Y$ in Fig.2(a) is the band inversion between Te$_d$ and Te$_z$ $p$ orbitals with a small gap of 30 meV giving a STI phase with the topological index[59] of (1;110). With broken IS induced by the second $B_{1u}$ mode (44.1 $cm^{-1}$) lattice distortion ($\lambda$=1.0) in Fig.2(b), the band double degeneracy along most high-symmetry directions are lifted. Especially along $\Gamma$-$Y$ direction, the top valence and bottom conduction band move closer toward each other and still maintain the band inversion as zoomed in Fig.2(c). This indicates WPs arising from the crossings between valence and conduction bands are nearby and slightly off the high-symmetry direction. After searching over the whole BZ with the MLWF-based TB model, we found two pairs of WPs near the $\Gamma$ point.

Although the lattice distortion is along the $c$-axis, the two pairs of WPs emerge on the $k_x$-$k_y$ plane at $k_z$=0. The Berry curvature has been calculated for one of two pairs of WPs along $k_x$ in Fig.2(d) and confirms these band crossings are indeed WPs with chirality or Chern number $C$=+/−1, i.e., source and sink of Berry curvature. Their energy-momentum coordinates are (E$_F$; ±0.002, ±0.076, 0.000 Å$^{-1}$). The other pair of WPs is related to this pair by the mirror or 2-fold rotation. The distance among WPs along $k_y$ is much larger than that along $k_x$. With increasing magnitude of the $B_{1u}$ mode lattice distortion, as shown in Fig.2(e) and (f) for $\lambda$=1.5 and 2.0, respectively, the band gap along $\Gamma$-$Y$ direction increases while maintaining the same band inversion features. The location of the band inversion along the $\Gamma$-$Y$ direction changes very little. the increasing size of the gap means the location of the WPs must move mostly in the $k_x$ direction. Indeed, we find the energy-momentum coordinates of the two pairs of WPs are moved to ($E_F$+4meV; ±0.010, ±0.089, 0.000 Å$^{-1}$) and ($E_F$+9meV; ±0.014, ±0.104, 0.000 Å$^{-1}$) for $\lambda$=1.5 and 2.0, respectively.

## C. (001) surface: expansive open Fermi arcs and connectivity changes

One unique behavior of WPs is the open Fermi arcs connecting the WP projections of opposite chirality on surfaces if there is no overlap. For ZrTe$_5$, this surface is (001), where all four WPs have their own projections (see Fig.1(b)). As shown in Fig.3(a) for (001) top surface for the second lowest $B_{1u}$ lattice distortion with $\lambda$=1.0, the projections of WPs form open Fermi arcs along $k_y$, while the distance along $k_x$ is much smaller. The (001) bottom surface (Fig.3(b)) have the same termination as the (001) top surface, but becomes non-equivalent because of the $B_{1u}$ mode breaking the IS. The open Fermi arcs on (001) bottom with $\lambda$=1.0 is more extensive than that on (001) top surface. When the magnitude of lattice distortion increases from $\lambda$=1.0 to 1.5, the distances among WPs increase mostly along $k_x$. As the WPs are pushed away from each other, the Fermi arcs are extended accordingly (Fig.3(c) and (d)). Notably, the open Fermi arcs also move closer to the other surface states. The open Fermi arcs are near the $\Gamma$ point and the connectivity of the Fermi arcs so far is still within the first SBZ or intra-SBZ.

However, when the magnitude of the $B_{1u}$ lattice distortion increase further to $\lambda$=2.0, we find that the open Fermi arcs connectivity changes from intra-SBZ to inter-SBZ. To clearly show the open Fermi arcs connectivity with respect to the (001) SBZ, the 2D Fermi surface (FS) for $\lambda$=2.0 at $E_F$+9 meV of the top surface is plotted with a zoom-out view in Fig.4(a). Strikingly although these WPs are still at general momentum near $\Gamma$ point, unlike the open Fermi arcs for smaller $\lambda$ or in other IS-broken WSMs, they extend beyond the first SBZ, cross the SBZ boundary and connect to those WP projections in the second SBZ along $k_y$. Such expansive open Fermi arcs on ZrTe$_5$ (001) surface is unique, as most open Fermi arcs connecting conventional WPs are small and intra-SBZ. Only open Fermi arcs connecting the highly degenerated Kramers-Weyl pinned at TRIM in chiral materials have such expansive range. On ZrTe$_5$ (001) bottom surface in Fig.4(g), the open Fermi arcs is expansive too, but it does not cross the SBZ boundary at $\lambda=2.0$.

To reveal the reason for such expansive open Fermi arcs in the induced $ZrTe_5$ WSM, we use the STI phase of the undistorted lattice as reference and inspect the topological surface states. For the STI phase of ZrTe5, unlike the easy cleaving (010) with only a single SDP at $\bar{\Gamma}$, (001) (Fig.4(b)) is known to have SDPs protected by TRS at three out of four TRIM points at $\bar{\Gamma}$, $\bar{S}$ and $\bar{S}'$. The last TRIM point at $\bar{Y}$ has a sizable band gap. The SDP at $\bar{\Gamma}$ is right at $E_F$, while the other two SDPs at $\bar{S}$ and $\bar{S}'$ have much lower energy at $E_F$–0.55 eV. Importantly these three SDPs are connected by the topological surface states expansive in the SBZ as seen in Fig.4(b). The topological surface states have large energy dispersion along most directions in the SBZ, except for $\bar{S}$-$\bar{S}'$, which is along $k_y$ or $b$-axis. This is understandable because $k_y$ direction is along the loosely vdW stacking direction. Such expansive surface states with similar band dispersion behavior also exist between bulk band $N$ and $N$-2, almost parallel in energy-momentum profile at slightly lower energy. For the 2D FS of STI in Fig.4(e) at $E_F$+9 meV, there is clearly a SDP at $\bar{\Gamma}$. Then there is a small hole pocket along $\bar{\Gamma}$-$\bar{X}$ direction. The other SDPs are shown by the topological surface states running mostly along $k_y$ direction and connecting to the neighboring SBZ. The surface states almost stay as straight lines along $k_y$, except around the $\bar{\Gamma}$ point with a bump.

With the second lowest $B_{1u}$ lattice distortion of $\lambda=2.0$ to remove IS, the SDP at $\bar{\Gamma}$ point is gapped out as shown in Fig.4(c) and (d) for the top and bottom surface, respectively. For the top surface, the SDP at $\bar{\Gamma}$ point breaks off from bulk conduction band and moves to lower energy. The surface bands along other directions also move to lower energy and become just surface valence states, except that the band along $\bar{S}$-$\bar{Y}$ still connects with bulk conduction band. As the result, for 2D FS in Fig.4(f), the SDP at $\bar{\Gamma}$ disappear and the nearby hole pockets shrinks for shifting to lower energy, while two pairs of WP projections emerge. Most noticeable, the surface states running along $k_y$ with a bump at $\bar{\Gamma}$ are now deformed to connect these four WP projections and become the expansive open Fermi arcs. In contrast for the bottom surface in Fig.4(d), the SDP at $\bar{\Gamma}$ breaks off from the bulk valence band and move to higher energy. The surface band along other directions also move to higher energy and become almost all surface conduction bands, again with the exception along the $\bar{S}$-$\bar{Y}$ direction for the one surface band still connecting to bulk valence band. Because the surface states move to higher energy, the change in 2D FS for bottom surface in Fig.4(g) is quite large. As the SDP at $\bar{\Gamma}$ disappear, the two small hole pockets enlarge into open Fermi arcs connecting the four WP projections, but they do not cross the SBZ boundary yet, unlike those on the (001) top surface. The surface states running along $k_y$ with bumps are pushed further away from $\bar{\Gamma}$ and become straight lines along $k_y$.

Thus, the topological surface states do not disappear upon topological phase transition to WSM, but deform into expansive open Fermi arcs. The spin texture in Fig.4(i) and (j) shows spin-momentum locking along the open Fermi arcs. Expansive open Fermi arcs have only been observed so far for Kramers-Weyl points pinned at TRIM in chiral crystals[17-20], such as RhSi and CoSi. For the two pairs of conventional WPs of $C=+/-1$ at general momentum here in $ZrTe_5$ induced by the second lowest $B_{1u}$ mode of 44.1 $cm^{-1}$, we predict expansive open Fermi arcs on the $ZrTe_5$ (001) surface. Note such expansive open Fermi arcs are topologically non-trivial features and also robust as shown in the Appendix for using PBEsol, a different XC functional in the DFT calculations. Extra calculations with the second lowest $B_{1u}$ lattice distortion imposed on both the theoretically optimized (a STI) and slightly volume-expanded (a WTI) structural parameters also give similar expansive open Fermi arcs for $ZrTe_5$, because the topological surface states are similarly expansive on (001) in these two phases, with the STI having three while the WTI having two SDPs[26].

Additionally, in Fig.5(a), we show the crossover of the open Fermi arcs with the topological surface states at $\lambda=1.8$ for the second lowest $B_{1u}$ lattice distortion that changes the Fermi arcs connectivity from intra-SBZ to inter-SBZ on the (001) top surface. The energy-momentum coordinates are ($E_F+7meV$; $\pm0.013, \pm0.098, 0.000$ Å$^{-1}$). When $\lambda$ increases further to 3.0, the open Fermi arcs connectivity on the (001) top surface remain inter-SBZ and expansive with increasing distance among the WPs (Fig.5(c)) for the energy-momentum coordinates of ($E_F+15meV$; $\pm0.019, \pm0.135, 0.000$ Å$^{-1}$). Now the Fermi arcs connectivity on the (001) bottom surface also changes from intra-SBZ (Fig.5(b)) to inter-SBZ (Fig.5(d)). Thus, not only this second lowest $B_{1u}$ mode in $ZrTe_5$ can be used to induce WPs, but also the magnitude of lattice distortion can be used to change the connectivity of the open Fermi arcs. We propose that coherent phonon of the second lowest $B_{1u}$ mode of 44.1 $cm^{-1}$ in optical pump-probe experiment can be used to dynamically change the surface Fermi arcs connectivity on $ZrTe_5$ (001) surface.

In a distinct contrast, for the first lowest $B_{1u}$ mode of 25.7 $cm^{-1}$ in ZrTe$_5$, the induced open Fermi arcs on (001) surface are not expansive as shown in Fig.6. With the first $B_{1u}$ lattice distortion increases from $\lambda$=2.0 (Fig.6(a) and (b)) to 6.0 (Fig.6(c) and (d)), the WPs separation increases along $k_x$, but the surface states have only small changes compared to the cases for the second lowest $B_{1u}$ lattice distortion. The connectivity of the open Fermi arcs induced by the first $B_{1u}$ mode does not change and remains only inside the first SBZ. The spectral functions in Fig.6(e) and (f) show also the much smaller change than those in Fig.4(c) and (d), which reflects in the much smaller energy shift of the SDPs at $\bar{S}$ and $\bar{S}'$ points. Notably for the second lowest $B_{1u}$ mode (Fig.1(f)), there is a large displacement of Te$_a$ along the $c$-axis in perpendicular to the (001) surface, which can induce a much larger change in (001) topological surface states than the first lowest $B_{1u}$ mode (Fig.1(e)). Thus, topological surface states connecting multiple SDPs provide the stage for expansive open Fermi arcs, but the lattice distortion to break the IS still needs to be sufficiently large enough to deform the surface states.

## D. (010) surface: surface states

Lastly, on ZrTe$_5$ (010) surface, in contrast to (001) surface, there are no expansive surface states for both STI and WSM in Fig.7(a) and (b), respectively. As zoomed in around $\bar{\Gamma}$ point, for STI in Fig.7(c), there is a single SDP at $\bar{\Gamma}$ protected by TRS inside the bulk band gap. With the second lowest $B_{1u}$ lattice distortion of $\lambda$=2.0, there are two pairs of WPs induced on the $k_z$=0 plane. While the SDP is gapped out, the bulk band structure becomes gapless, as shown in Fig.7(d) by the touching point between bulk valence and conduction band projections along $\bar{\Gamma}$-$\bar{U}$. There is a short segment of surface state along $\bar{\Gamma}$-$\bar{U}$ coming out from the projections of the two pairs of WPs. As for the 2D FS on (010) surface at E$_F$+9 $meV$ in the STI (Fig.7(e)), there is a circle arising from the surface Dirac cone around the $\bar{\Gamma}$ point. With the WSM induced by the $B_{1u}$ lattice distortion (Fig.7(f)), the circle from the surface Dirac cone disappear. Although there are no open Fermi arcs because WPs of opposite chirality are projected onto each other, there are surface states at WP projections, just not open Fermi arcs.

As a final note, although the IS is broken by these $B_{1u}$ modes, the induced two pairs of WPs are at the same energy because of the remaining mirror symmetry. Further symmetry-breaking is needed to separate the WPs to different energies to realize the recently proposed quantized circular photo-galvanic effect[11].

## IV. Conclusion

In conclusion, we have shown that low-frequency infrared phonon modes in ZrTe$_5$ can be used to induce Weyl points by breaking inversion symmetry while still maintaining band inversion near $\Gamma$ point. The two pairs of Weyl points induced by the second lowest $B_{1u}$ mode beyond certain lattice distortion are connected by expansive open Fermi arcs crossing into neighboring surface Brillouin zones. Our study indicates that expansive open Fermi arcs are not unique to unconventional chiral fermions or Kramers-Weyl points pinned at time-reversal invariant momentum in chiral materials, but can also exist for conventional Weyl points with chirality of +/–1 at general momentum. The key for such expansive open Fermi arcs is the transformation of the topological surface states connecting the multiple surface Dirac points on (001) surface of ZrTe$_5$ upon topological phase transition. Furthermore, we find that the connectivity of the open Fermi arcs can be changed with different magnitude of the second lowest $B_{1u}$ lattice distortion. Thus, we propose that topological surface states connecting multiple surface Dirac points with large enough lattice distortion from infrared phonon modes are platforms to induce expansive open Fermi arcs. Our study also shows the potential of using coherent optical control and ultrafast spectroscopy to induce and detect novel topological features including expansive open Fermi arcs and dynamically control of Fermi arcs connectivity in ZrTe$_5$.

## Acknowledgements

We acknowledge helpful discussion with Jigang Wang. This work was supported by the Center for the Advancement of Topological Semimetals, an Energy Frontier Research Center funded by U.S. Department of Energy (DOE), Office of Science, Basic Energy Sciences. Work was performed at Ames Laboratory, which is operated by Iowa State University under contract DE-AC02-07CH11358.

| Mode ($cm^{-1}$) | LDA | PBE-D2 | Expt |
|---|---|---|---|
| $B_{1g}$ | 39.2 | 36.7 | 28 |
| $A_{1g}$ | 43.4 | 37.6 | 39 |
| $B_{3g}$ | 48.7 | 41.8 | 48 |
| $B_{3g}$ | 80.1 | 79.4 | 62 |
| $B_{2g}$ | 82.8 | 72.7 | 72 |
| $B_{2g}$ | 99.0 | 96.3 | 88 |
| $B_{1u}$ | 20.0 | 25.7 | |
| $B_{1u}$ | 62.0 | 44.1 | |

Table 1. Zone-center optical phonon modes calculated in LDA and PBE-D2 in comparison to experimental data[56].

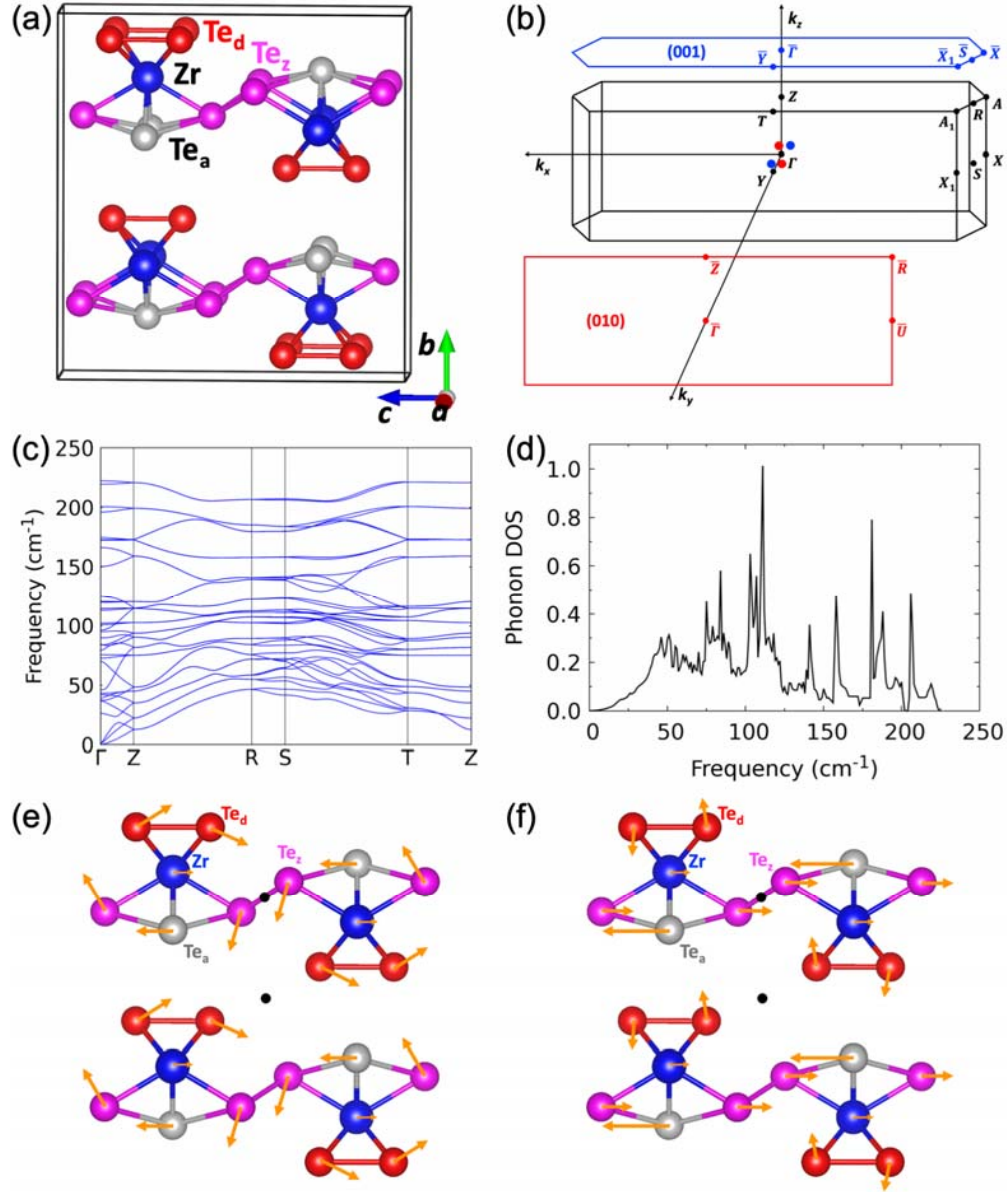

Figure 1. (a) Base-centered orthorhombic structure of ZrTe$_5$ in space group 63 (*Cmcm*) with different sites of Zr, apical Te (Te$_a$), dimer Te (Te$_d$) and zigzag Te (Te$_z$) labeled. (b) Bulk, (001) and (010) surface Brillouin zones labeled with high symmetry points. Large red and blue circles are the two pairs of Weyl points induced with $B_{1u}$ lattice distortion. (c) Calculated phonon dispersion and (d) phonon density of states (DOS). (e) First and (f) second lowest $B_{1u}$ infrared phonon mode with lattice distortion shown by arrows, respectively. Black circles show the broken inversion centers.

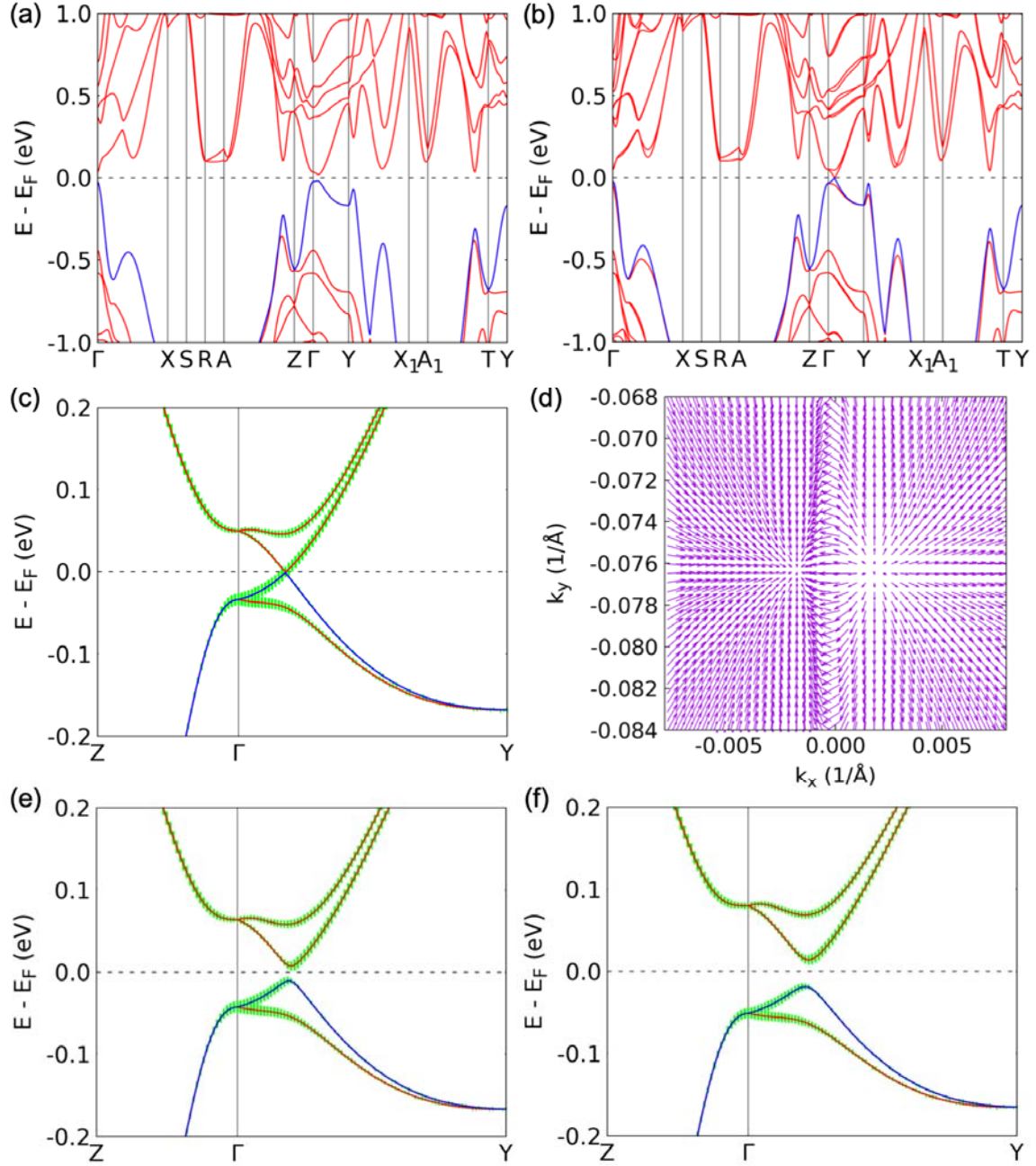

Figure 2. Band structures of ZrTe$_5$ in experimental lattice parameters without (a) and with (b) the second lowest $B_{1u}$ (44.1 $cm^{-1}$) lattice distortion of magnitude $\lambda$=1.0. (c) Band structure of (b) zoomed in along $Z$-$\Gamma$-$Y$ with projection (green shadow) on Te$_d$ $p$ orbitals. The top valence band is shown in blue. (d) Source and sink of Berry curvature on $k_x$-$k_y$ plane at $k_z$=0.0 showing a pair of Weyl points near $\Gamma$ as labeled in Fig.1(e). Panel (e) and (f) are the same as (c) with the increased magnitude of $B_{1u}$ lattice distortion at $\lambda$=1.5 and $\lambda$=2.0.

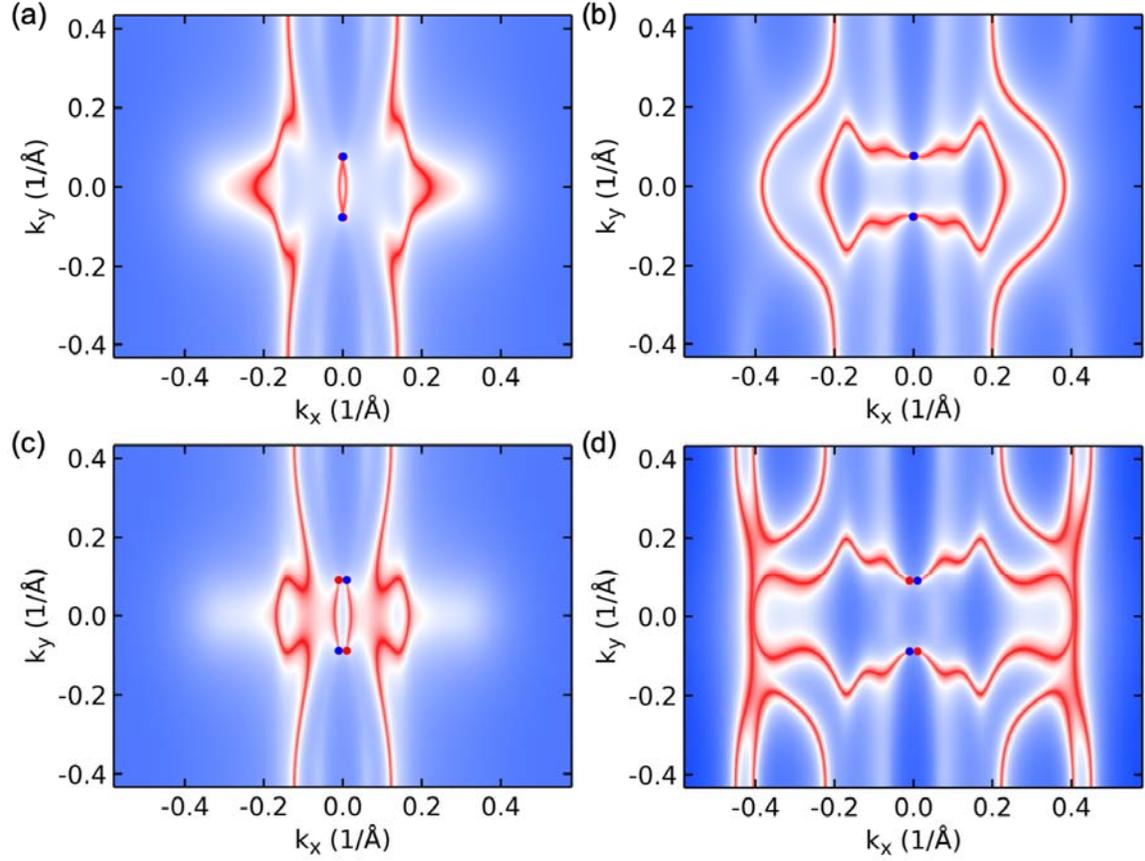

Figure 3. ZrTe$_5$ (001) surface 2D Fermi surface (FS) with the second lowest $B_{1u}$ lattice distortion of different magnitude $\lambda=1.0$ ((a) and (b)) and $\lambda=1.5$ ((c) and (d)) at Weyl point binding energy. Panel (a) and (c) are the (001) top, and (b) and (d) are the (001) bottom surfaces. Low, medium, and high density of states is indicated by blue, white and red colors, respectively. Red (blue) circles label the Weyl point projection with Chern number $C=+1(-1)$.

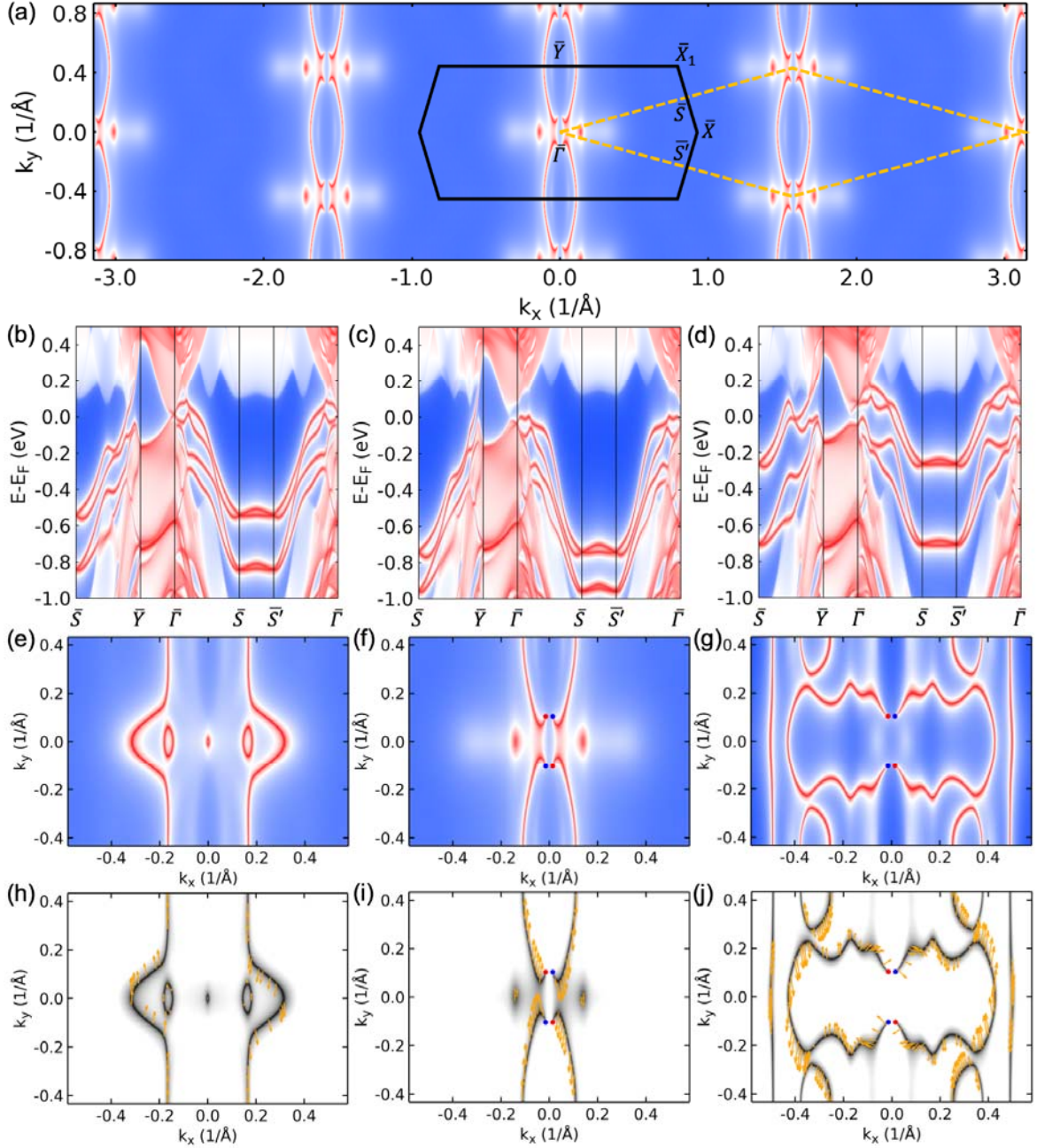

Figure 4. ZrTe$_5$ (001) surface states and 2D FS at E$_F$+9 meV. (a) 2D FS with the second lowest $B_{1u}$ lattice distortion of magnitude $\lambda$=2.0 zoomed out to show the surface reciprocal unit cell (orange) and Brillouin zone (black) with high symmetry points labeled. Panel (b), (e) and (h) are the surface states, 2D FS and spin texture, respectively, on (001) surface of the STI without lattice distortion. Panel (c), (f) and (i) are those on the (001) top surface of the WSM with $B_{1u}$ lattice distortion. Panel (d), (g) and (j) are those on the (001) bottom surface of the WSM with $B_{1u}$ lattice distortion. Low, medium, and high density of states is indicated by blue, white and red colors, respectively. In (f), (g), (i) and (j), red (blue) circles label the Weyl point projection with Chern number $C$=+1(−1).

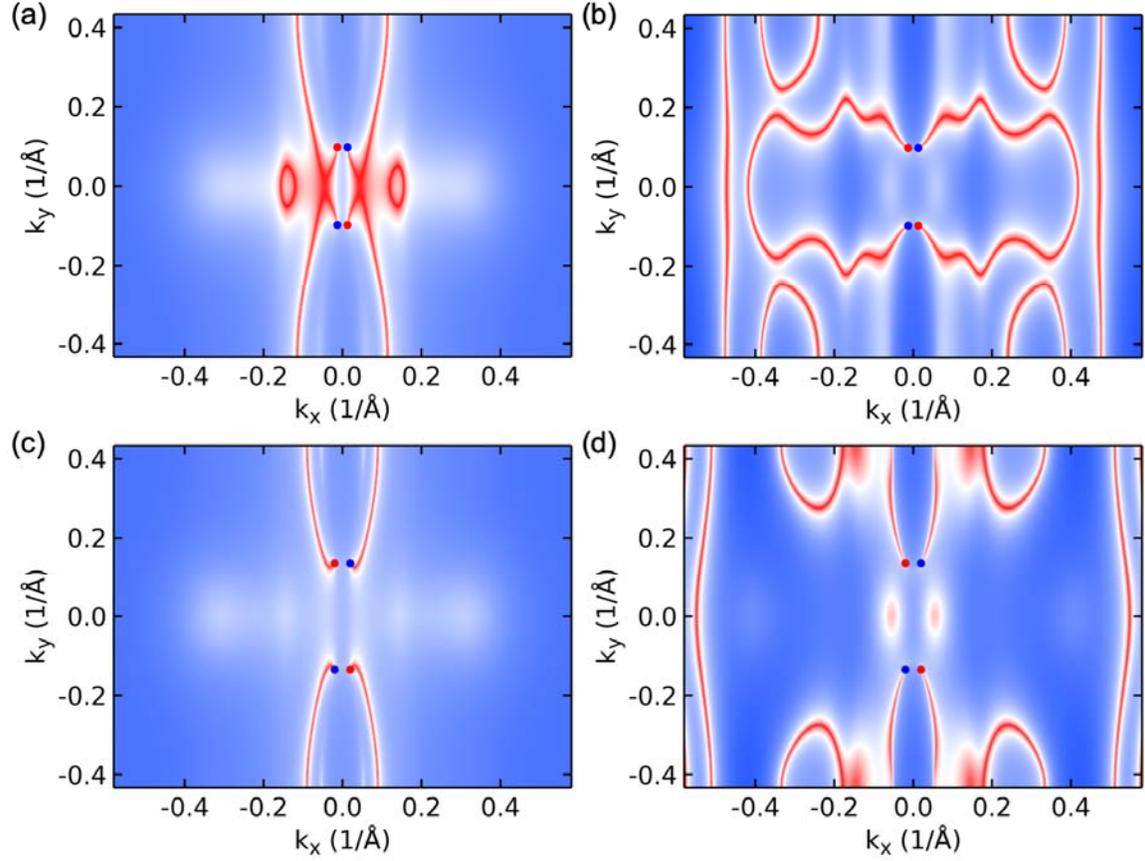

Figure 5. ZrTe$_5$ (001) surface 2D FS with the second lowest $B_{1u}$ lattice distortion of different magnitude $\lambda=1.8$ ((a) and (b)) and $\lambda=3.0$ ((c) and (d)) at Weyl point binding energy. Panel (a) and (c) are the (001) top, and (b) and (d) are the (001) bottom surfaces. Low, medium, and high density of states is indicated by blue, white and red colors, respectively. Red (blue) circles label the Weyl point projection with Chern number $C=+1(-1)$.

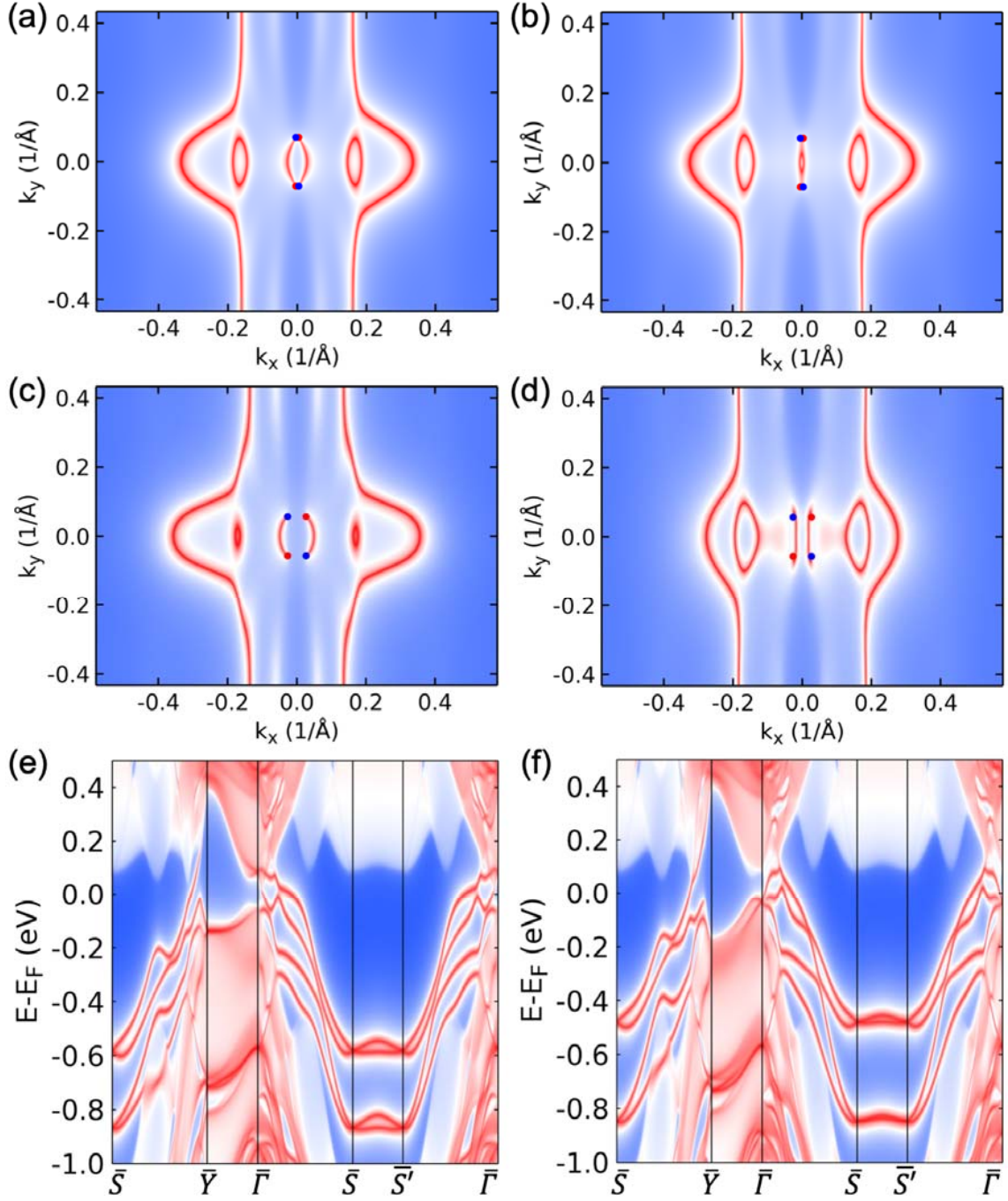

Figure 6. ZrTe$_5$ (001) surface 2D FS with the first lowest $B_{1u}$ lattice distortion of different magnitude $\lambda=2.0$ ((a) and (b)) and $\lambda=6.0$ ((c) and (d)) at Weyl point binding energy. (e) and (f) The surface spectral functions for $\lambda=6.0$. Panel (a), (c) and (e) are the (001) top, and (b), (d) and (f) are the (001) bottom surfaces. Low, medium, and high density of states is indicated by blue, white and red colors, respectively. Red (blue) circles label the Weyl point projection with Chern number $C=+1(-1)$.

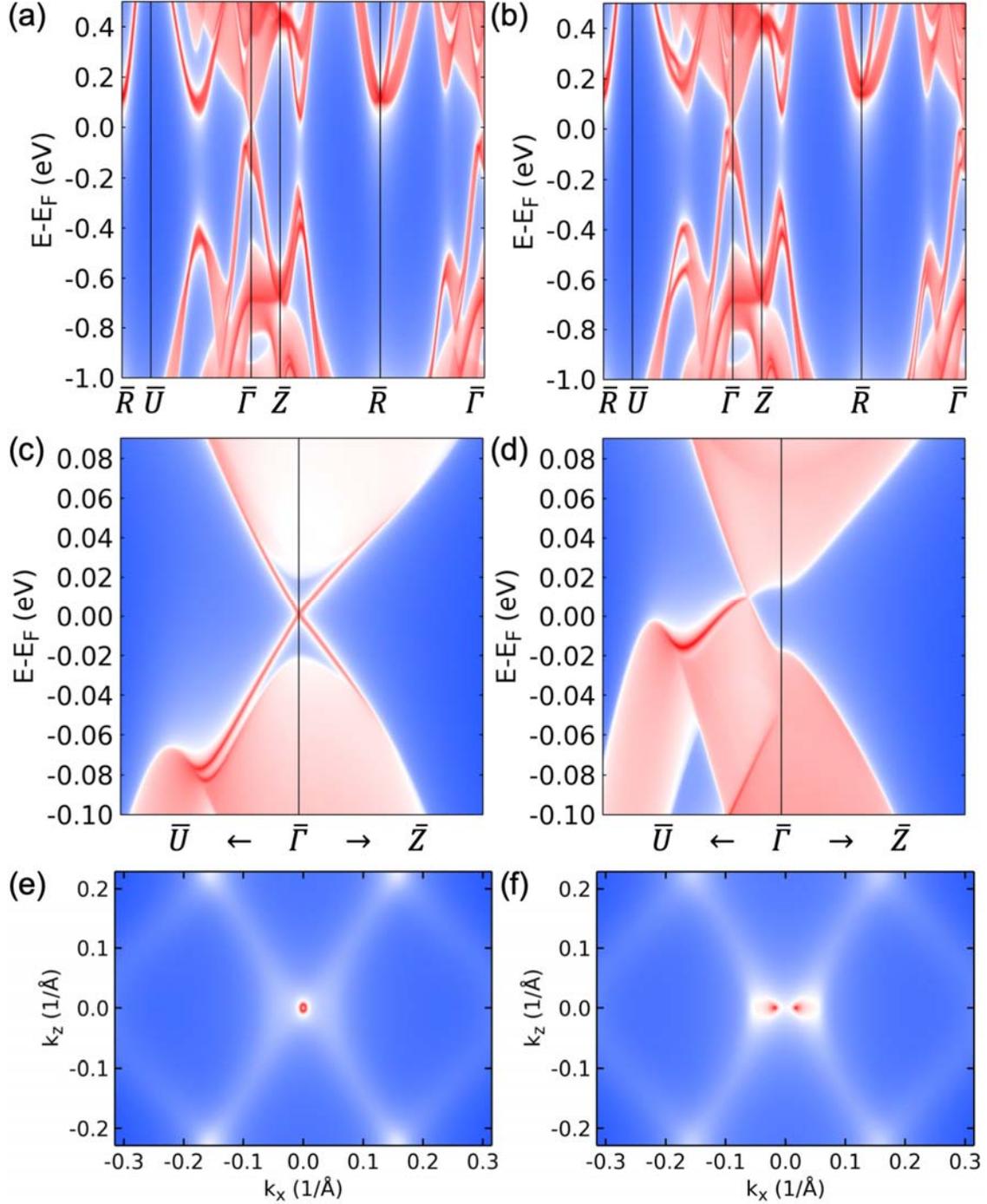

Figure 7. ZrTe$_5$ (010) surface states and 2D FS at E$_F$+9 meV. Panel (a), (c) and (e) are the surface states, zoom-in, 2D FS of the STI, respectively, without lattice distortion. Panel (b), (d) and (f) are those for WSM with the second lowest $B_{1u}$ lattice distortion of magnitude $\lambda$=2.0. Low, medium, and high density of states is indicated by blue, white and red colors, respectively. In (f), a pair of Weyl points with opposite chirality are projected on top of each other.